\documentclass[aps,twocolumn,showpacs]{revtex4}
\usepackage{graphics}
\begin{document}
%\baselineskip=.4in
%\normalsize
\title{\Large{Ground state phase diagram and magnetoconductance of a one dimensional Hubbard superlattice at half-filling}}
\author{Jayeeta Chowdhury$^1$}
\author{S. N. Karmakar$^2$}
\author{Bibhas Bhattacharyya$^3$}
\affiliation{$^1$Department of Physics, East Calcutta Girls' College, Lake Town, Lake Town Link Road, Kolkata 700 089, India\\
$^2$TCMP Division, Saha Institute of Nuclear Physics, 1/AF Bidhannagar, Kolkata 700 064, India\\
$^3$Department of Physics, Scottish Church College, 1 \& 3 Urquhart Square, Kolkata 700 006, India}
\begin{abstract}
We have studied a one dimensional Hubbard superlattice with different Coulomb correlations at alternating sites for a 
half-filled band.  Mean field calculations based on the Hartree-Fock approximation together with a real space renormalization
group technique were used to 
study the ground state of the system. The phase diagrams obtained in these approaches agree with each other  from the weak 
to the  
intermediate coupling regime. The mean field results show very quick convergence with system size. The renormalization group 
results indicate a spatial modulation of local moments that was identified in some previous work. Also we have studied the 
magnetoconductance of such superlattices which reveals several interesting points.
\end{abstract}
\maketitle
\section{Introduction}
The study of electronic correlation remains at the focus of recent theoretical interest due to the development of different novel materials of which the metallic multilayers \cite{Heinrich} form a very important component. The oscillation of exchange coupling between magnetic layers \cite{Grunberg} and the appearance of giant magnetoresistance \cite{Baibich} are among the exciting features of the magnetic multilayers, e.g. the layered Fe/Cr structures. An intense theoretical attempt has been made to understand the magnetic behavior of such systems \cite{Santos1,Santos2,Santos3,Santos4,Santos5}. In these pioneering works in this field a simple generalization of the one dimensional Hubbard model was proposed to investigate the role of electronic correlation in one dimensional superlattices. This model consists of a periodic arrangement of $N_U$ sites in which the on-site Coulomb correlation $(U)$ is repulsive followed by $N_0$ sites with no on-site interaction $(U=0)$. It was found that such a model gives rise
  to some interesting features, in sharp contrast with the magnetic behavior observed in an otherwise homogeneous system (conventional one band Hubbard model) \cite{Santos1}.

However, most of the previous works in this model employed exact diagonalization of finite systems typically having $8-24$ sites \cite{Santos1, Santos2}. Some of these works reveal preferential distribution of local moments on sublattices and observe the suppression of spin density wave (SDW) order \cite{Santos1}. Some others deal with problems like metal-insulator transition (MIT) \cite{Santos2} or formation of charge density wave \cite{Santos3} in such systems. Density matrix renormalization group (DMRG) calculations are also performed to overcome limitations on system size imposed by the technique of exact diagonalization; and, these consider system sizes $\sim 48-150$ \cite{Santos4}. Very recently a generalization of the said model has been considered where, instead of two types of sites with on-site correlation parameters $U>0$ and $U=0$ respectively, one takes into account two different values of $U(>0)$ at adjacent sites \cite{Japaridze}; however, a detailed study of this
generalized model is yet to be worked out. In view of the wider applicability
  of this model to diverse experimental systems, we consider here a preliminary study of this alternating Hubbard model 
(AHM) in one dimension. Also we aim at observing the behavior of the same when the system size is reasonably large.

In the present study, we investigate the ground state properties of the AHM in one dimension and for a half-filled band by using a Hartree-Fock approximation (HFA) together with a real space renormalization group (RG) calculation. These two techniques, complemented by one another, was found to be very successful in studying similar cases in recent past \cite{Gupta1}. Apart from constructing the ground state phase diagram we also investigate the magnetoconductance of a finite chain  within the HFA scenario. In sec.II we introduce the model and give the HFA calculations. Sec.III describes the results obtained in HFA. Sec.IV contains some details of the RG scheme while sec.V shows the RG results together with a comparison between the same and that obtained in the HFA. In sec.VI we present 
the results on magnetoconductance of the model and sec.VII summarizes the 
present work.
\section{The model and the Hartree-Fock calculations}
Our model is defined on a  one dimensional Hubbard chain of $N$ (even integer) sites consisting of two sublattices. 
The model Hamiltonian is,
\begin{eqnarray}
& H&=t\sum_{i,\sigma}(c_{i,\sigma}^{\dag}c_{i+1,\sigma}+h.c.)\nonumber\\
&&+U_{A}\sum_{i\in \cal{A}}n_{i,\uparrow}n_{i,\downarrow}+U_{B}\sum_{i\in \cal{B}}n_{i,\uparrow}n_{i,\downarrow}~~~~~~
\end{eqnarray}

\noindent 
The two sublattices constructed out of odd and even numbered sites are labeled 
by $\mathcal{A}$ and $\mathcal{B}$ respectively.
$c_{i,\sigma}^{\dag}(c_{i,\sigma})$ is the creation(annihilation) operator for 
an electron with spin $\sigma$ at the $i$-th site. 
$n_{i,\sigma}=c_{i,\sigma}^{\dag}c_{i,\sigma}$, and $n_{i}=\sum_{\sigma}n_{i,\sigma}$
 is the number operator at the $i$-th site.
$t$ is the hopping integral between nearest neighbor sites.
%~$\epsilon_{A}$, $\epsilon_{B}$ are the site energies and 
$U_{A}$, $U_{B}$ are the on-site Coulomb repulsion energies on the sites corresponding to two sublattices $\mathcal{A}$ and $\mathcal{B}$ respectively.

We decouple the Hamiltonian within  the HFA, which is expected to work at least in the weak coupling regime. 
We define two parameters, number of electron  $N_{i}$ and the magnetization $M_{i}$ at the $i$-th site, where, 
\begin{eqnarray}
&&N_{i}=\langle n_{i,\uparrow}\rangle +\langle n_{i,\downarrow}\rangle\nonumber\\
&&M_{i}=\langle n_{i,\uparrow}\rangle -\langle n_{i,\downarrow}\rangle
\end{eqnarray}
%
%\noindent
%Using the Hartree-Fock Approximation (HFA) treatment for decoupling
%\begin{eqnarray}
%\sum_{i}n_{i,\uparrow}n_{i,\downarrow}=\langle n_{i,\uparrow}\rangle n_{i,\downarrow}+n_{i,\uparrow}\langle n_{i,\downarrow}\rangle&&\nonumber\\
%-\langle n_{i,\uparrow}\rangle \langle n_{i,\downarrow}\rangle&&
%\end{eqnarray}

%\noindent
These lead to a decoupled Hamiltonian:
\begin{eqnarray}
&&H= t\sum_{i,\sigma}(c_{i,\sigma}^{\dag}c_{i+1,\sigma}+h.c.)\nonumber\\
&&+\frac{U_{A}}{2}\sum_{i\in {\cal A}}[(N_{i}+M_{i})n_{i,\downarrow}+(N_{i}-M_{i})n_{i,\uparrow})]\nonumber\\
&&+\frac{U_{B}}{2}\sum_{i\in {\cal B}}[(N_{i}+M_{i})n_{i,\downarrow}+(N_{i}-M_{i})n_{i,\uparrow})]\nonumber\\
&&-\frac{1}{4}(U_{A}+U_{B})(N_{i}^{2}-M_{i}^{2})
\end{eqnarray}

\noindent
Now the Hamiltonian can be divided into two parts for two types of spins, i.e.~$H=H_{\uparrow}+H_{\downarrow}$. 
%and each part can now be easily diagonalized 
In an unrestricted HFA one diagonalizes $H_{\uparrow}$ and $H_{\downarrow}$ in a self-consistent manner to 
obtain the single particle energy levels. The ground state can be constructed by filling up the energy levels from both the up and the down bands upto the Fermi level.

One can define the spin and the charge density order parameters, $c$ and $s$ respectively, by
$$ 
c= \frac{1}{N} \sum_i (-1)^i (n_{i,\uparrow}+n_{i,\downarrow}),\quad s=\frac{1}{N} \sum_{i} (-1)^i 
(n_{i,\uparrow} - n_{i, \downarrow}.
$$
We consider a  half-filled  chain with periodic boundary condition. 
One can easily check by using an unrestricted HFA calculation that all sites 
corresponding to a given sublattice become equivalent under a periodic boundary condition.
This leads to a simplification of the formulas for $c$ and $s$ given above.
%It was observed that all sites corresponding to a sublattice are equivalent. To draw the phase diagram, 
We, therefore, use these simplified forms of the charge density order parameter $c$ and the spin density order parameter or 
antiferromagnetic order parameter $s$ as given by,
\begin{eqnarray}
&&c=\frac{1}{2}\langle n_{B,\uparrow}+n_{B,\downarrow}-n_{A,\uparrow}-n_{A,\downarrow}\rangle\nonumber\\
&&s=\frac{1}{2}\langle n_{B,\uparrow}-n_{B,\downarrow}-n_{A,\uparrow}+n_{A,\downarrow}\rangle~~~
\end{eqnarray}

\noindent
It is to be noted here that $|c|=1$ for a perfect ``chess-board''-type CDW and 
$|s|=1$ for a perfect N\'{e}el-type antiferromagnetic SDW state.
We now investigate the dependence of these order parameters on the values of 
$U_{A}$ and $U_{B}$. 

\begin{figure}
\includegraphics*{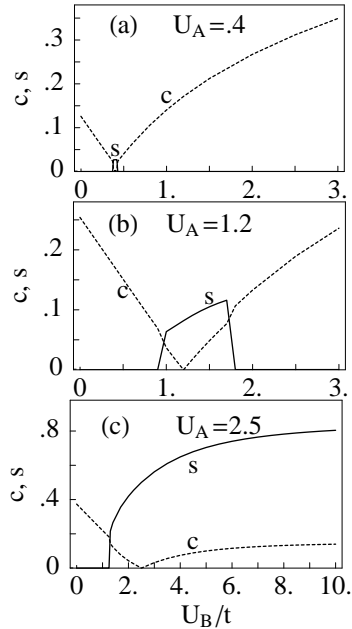}
\caption{Plot of the order parameters $c$ and $s$ as functions of $U_B/t$ for (a) $U_A =0.4$, (b) $U_A = 1.2$, and (c) $U_A=2.5$ (scale of energy is chosen by setting $t=1.0$) for $N =100$. The dotted line corresponds to the charge order parameter $c$ while the solid line shows the spin order parameter $s$ as calculated from HFA.}
\end{figure}

\section{The results of Hartree-Fock calculations}
In fig.1 we plot the order parameters $c$ and $s$ as functions of $U_B/t$ for 
different fixed values of $U_A/t$ (for a chain having sites $N = 100$).
Keeping $U_{A}/t$ at a fixed value ($>0$) and varying $U_{B}/t$ from $0$ to higher values, we find that initially the system is charge ordered. The electrons tend to
localize at the sites with lower Coulomb repulsion energies, keeping the other sites vacant. 
As a result a charge density wave (CDW) is formed. In this regime, the charge density order parameter $c$ assumes a high 
value while the spin density order parameter $s$ is zero. As we gradually increase $U_{B}/t$ keeping $U_{A}/t$ at the 
same fixed value, we see that there is a gradual fall in the value  of $c$. At a certain value of $U_{B}/t$  
there occurs a sharp rise in $s$ which now takes over the value of $c$. For small values of $U_A/t (<<1)$ the transition 
occurs at the point of homogeneity, i.e. at $U_B = U_A$. In this case, however, a further increase in $U_B/t$ 
suppresses the spin order again, and the CDW sets in. Therefore, the SDW is found to form only at a singular point $U_A=U_B$ 
which is in agreement with the known result for the ``homogeneous'' limit \cite{Shiba}. The situation becomes different 
for larger values of $U_A/t$. For an intermediate value of $U_A/t ( \sim 1)$, we find two transitions: one from a 
CDW to an SDW and then  from the SDW to a CDW again; this can be identified by two successive cross-overs in the $c$- and 
the $s$-curves (fig.1b). It is interesting to note that the charge order vanishes only at the point of homogeneity. For 
large $U_A/t (>> 1)$ there appears only one transition from a CDW state to an SDW state at a specific value of $U_B/t (<
U_A/t)$. Here also, the charge order parameter vanishes at $U_A = U_B$, and then rises slowly with $U_B/t (> U_A/t)$. However, in 
this region, the spin order parameter always dominates over the charge order parameter (fig.1c). 
The phase transitions occurring at points of cross-overs of $c$ and $s$ can further be explored by studying the gap in the 
spectrum at the Fermi level. The energy gap $\Delta_{HFA}$ at the Fermi level of a system containing $n$ electrons can be 
estimated from
$$
\Delta_{HFA} = E_{n+1}-E_n
$$
where, $E_n$ is the ground state energy of (3) for a system of $n$ particles. In fig.2 we plot the energy gap $\Delta_{HFA}$
as a function of $U_B/t$ for different fixed values of $U_A/t$ at half-filling. Dips in the curves of $\Delta_{HFA}$ match 
with the 
corresponding values of $U_B/t$ that were identified as points of phase transitions in fig.1. It is to be noted here that
for large values of $U_A/t$ the energy gap sharply increases after the CDW/SDW transition has occurred (fig.2). In 
this regime the slow increase of the charge order parameter does not modify this behavior. 

\begin{figure}
\includegraphics*{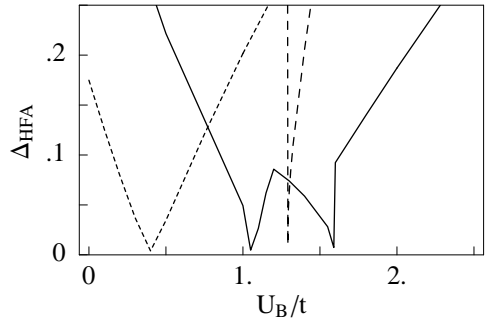}
\caption{Plot of the energy gap $\Delta_{HFA}$, as found from HFA, as a function of $U_B/t$ for $N=300$. We have used the dotted line for $U_{A}=0.4$, the solid line for $U_{A}=1.2$, and the dashed line for $U_{A}=2.5$ (scale of energy is chosen by setting $t=1.0$). Note that the zeroes in the gap match with the transition points identified from fig. 1.}
\end{figure}

Identification of the points of phase transitions by the methods mentioned above enables us to draw the phase diagram
of the model (1) on the $U_A/t - U_B/t$ plane within the HFA. 

\begin{figure}
\includegraphics*{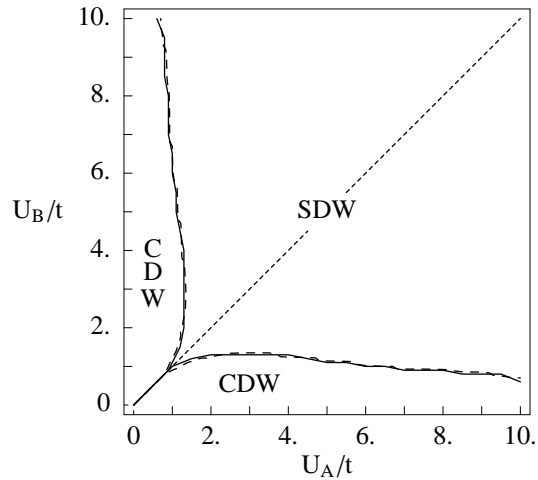}
\caption{The HFA phase diagram of the AHM for different system sizes (the dashed line for $N=100$, and the solid line for $N=200$) in the $U_A/t - U_B/t$ plane. The plot is symmetrical about the ``line of homogeneity'' $U_A=U_B$ which is shown by the dotted line.}
\end{figure}

In fig.3 we plot the phase diagrams for different $N$ values
to note the quick convergence of the present mean field results with system size. 
For small values of $U_{A}$ and $U_{B}$ the antiferromagnetic phase  actually occurs along the  line of ``homogeneity'' 
($U_{A}=U_{B}$). For higher values of $U_{A}$ and $U_{B}$ we obtain a centrally located broad SDW region together with 
two charge ordered phases located near the axes.  The phase diagram turns out to be 
perfectly symmetric about the line of homogeneity  along which the system is already known to be 
antiferromagnetic from exact calculations \cite{Shiba}. In the limit $U_B >> U_A$ ($U_A >> U_B$) the CDW/SDW transition line 
above (below) the line of homogeneity bends towards the $U_B/t$ ($U_A/t$) axis.
\section{The Renormalization Group calculations}
Next we apply a real space Renormalization Group (RG) technique \cite{Hirsch} to the same model keeping in view the success
of this technique to similar 1-d systems of correlated electrons \cite{Gupta1, Bhatta1, Bhatta2} in recent past. It is to be noted 
here that in order to achieve a closed parameter space under the present scheme of RG iteration, we must generalize the model
(1) as follows:
\begin{eqnarray}
& H&= \epsilon_A \sum_{i \in \cal{A}} n_i + \epsilon_B \sum_{i \in \cal{B}} n_i \nonumber \\
&&+t\sum_{i,\sigma}(c_{i,\sigma}^{\dag}c_{i+1,\sigma}+h.c.)\nonumber\\
&&+U_{A}\sum_{i\in \cal{A}}n_{i,\uparrow}n_{i,\downarrow}+U_{B}\sum_{i\in \cal{B}}n_{i,\uparrow}n_{i,\downarrow}~,~~~~~
\end{eqnarray}
where, $\epsilon_A$ ($\epsilon_B$) refers to the site energy of a site belonging to the $\cal{A}$ ($\cal{B}$) sublattice. We start
with $\epsilon_A = \epsilon_B =0$ which makes (5) equivalent to (1). However, in course of the RG iteration 
non-zero values of $\epsilon_A$ and $\epsilon_B$ may appear.

For implementing the RG transformation the whole chain is now divided into cells containing three sites each. Since our system is a 
bipartite 
lattice with two types of sites $A$ and $B$, there will appear two types of cells $ABA$ and $BAB$. At the renormalized 
length scale, we identify the $ABA$ cells as the new (renormalized) $A$-type sites and the $BAB$ cells as the new $B$-type 
sites. There are four different `on-site' states for each site 
$|0\rangle,|+\rangle,|-\rangle,|+-\rangle$.
We diagonalize the cell-Hamiltonian (for both types of cells), and among the eigenstates of the cell-Hamiltonian we 
retain only four low-lying states at each iteration, for construction of the RG recursion relations. We are interested in 
the half-filled ground state; so we 
retain the lowest energy states in the subspaces $\{n=2,S=S_{z}=0\}$, $\{n=3,S=\frac{1}{2},S_{z}=\pm{\frac{1}{2}}\}$ and 
$\{n=4,S=S_{z}=0\}$ of each type of cells. Here $n, S,$ and $S_z$ denote the total number of electrons, the total spin, and
the $z$-component of the total spin respectively. These four states in a cell can now be identified as the `renormalized' on-site states 
$|0'\rangle$,~$|+'\rangle$, ~$|-'\rangle$ and $|+-'\rangle$ respectively. 

To find the renormalized hopping matrix element, the matrix elements of $c_{\sigma}^{b}(A)$ and  $c_{\sigma}^{b}(B)$ between `renormalized on-site states' are calculated, where $c_{\sigma}^{b}$ is the annihilation operator of the electron with spin $\sigma$ at the boundary site of the cell and $A$ or $B$ in the parenthesis denotes the type of the cell. Let for an $ABA$ type cell 
\begin{eqnarray}
\langle 0'|c_{\uparrow}^{b}(A)|+'\rangle=\lambda_{1}(A)~,&&\nonumber\\
\langle -'|c_{\uparrow}^{b}(A)|+-'\rangle=\lambda_{2}(A)~,&&
\end{eqnarray}

\noindent
and for a $BAB$ type cell
\begin{eqnarray}
\langle 0'|c_{\uparrow}^{b}(B)|+'\rangle=\lambda_{1}(B)~,&&\nonumber\\
\langle -'|c_{\uparrow}^{b}(B)|+-'\rangle=\lambda_{2}(B)~.&&
\end{eqnarray}

\noindent
Our system possesses spin reversal symmetry, so the matrix elements for 
$c_{\downarrow}^{b}$'s will be as same as that of $c_{\uparrow}^{b}$'s (except
for a fermionic sign change in the value of $\lambda_2$). But due to the lack of particle-hole symmetry, $\lambda_{1}(A)\neq\lambda_{2}(A)$, and 
$\lambda_{1}(B)\neq\lambda_{2}(B)$. 
%All $\lambda$'s are constructed out of the elements of the eigenvectors corresponding to the ground states. We define
At this stage we introduce an approximation \cite{Gupta1, Bhatta1, Ma} by defining.
\begin{eqnarray}
&&\lambda(A)=\sqrt{\lambda_{1}(A)\lambda_{2}(A)}\nonumber\\
&&\lambda(B)=\sqrt{\lambda_{1}(B)\lambda_{2}(B)}
\end{eqnarray}

\noindent
which leads to
\begin{equation}
c_{\sigma}^{b}(\Gamma)=\lambda(\Gamma)c_{\sigma}'(\Gamma)
\end{equation}

\noindent
where $\Gamma=A,B$ and $\sigma=\uparrow,\downarrow$. So the effective renormalized hopping becomes
\begin{equation}
t'=\lambda(A)\lambda(B)t
\end{equation}

The intra-cell Hamiltonian, restricted to the subspace of the four states 
$|0'\rangle$, $|+'\rangle$, $|-'\rangle$ and $|+-'\rangle$, can now be written 
in terms of the new cell-fermion operators \cite{Hirsch} as
\begin{eqnarray}
&&H'=E_{0'}+(E_{+'}-E_{0'})(n_{\uparrow}'+n_{\downarrow}')\nonumber\\
&&+(E_{+-'}+E_{0'}-2E_{+'})n_{\uparrow}' n_{\downarrow}'
\end{eqnarray}

\noindent
where $E_{+-'}$, $E_{+'}$ and $E_{0'}$ are the lowest energies of the subspaces corresponding to 4, 3 and 2 particles respectively. From this we can easily identify the renormalized on-site quantities
\begin{equation}
U_{\Gamma}'=E_{+-'}(\Gamma)+E_{0'}(\Gamma)-2E_{+'}(\Gamma)
\end{equation}

\noindent
and
\begin{equation}
\epsilon_{\Gamma}'=E_{+'}(\Gamma)-E_{0'}(\Gamma)
\end{equation}

\noindent
where $\Gamma=A,B$.

\noindent
The ground state energy per site is computed from the sum
\begin{equation}
E_{0}=\frac{1}{2}\sum_{n=1}^{\infty}\frac{[E_{0'}^{(n)}(A)+E_{0'}^{(n)}(B)]}{3^n}
\end{equation}

\noindent
where $n$ denotes $n$ th stage of iteration. We also calculate the local moment $L_0$ defined by
$$
L_0 = \frac{3}{4} \left(n_{\uparrow} - n_{\downarrow} \right)^2~.
$$
In absence of particle-hole symmetry this leads to a recursion relation of the 
form \cite{Gupta1}
\begin{equation}
L_0 = a + b L'_0 + c P'~,
\end{equation}
where, $P= \left(n_{\uparrow}+n_{\downarrow}\right) \left( n_{\uparrow} +
n_{\downarrow} -1 \right)$ obeys a similar recursion relation given by
\begin{equation}
P= d + e L'_0 + f P'~,
\end{equation}
with $a, b, c, d, e,$ and $f$ are obtained from the matrix elements of $L_0$ and $P$ between the truncated basis of the cell Hamiltonian. The operators $L_0$ and $P$ are considered for the central site of the cell to minimize the boundary effects \cite{Hirsch, Gupta1}. However, in implementing the recursion relation it is to be noted that a recursion for $L_0$ at a $B$ type site (being at the 
middle of an $ABA$ cell) will involve $L'_0$ and $P'$ pertaining to a renormalized $A$ type site, because under the RG 
transformation the $ABA$ cell $\rightarrow$ a `renormalized' $A$ site. Similar consideration arises for $P$ as well. $L_0$ found for two different types of sites need not be equal to each other, in general. 

To see the nature of the short range spin correlation we further compute the nearest neighbor (nn) spin correlation function
$\langle S_{1z} S_{2z} \rangle$, and the next nearest neighbor (nnn) spin correlation function $\langle S_{1z} S_{3z} \rangle$
in each type of cells, where $S_{iz} = n_{i,\uparrow} -n_{i,\downarrow}$. Recursion relations are very much similar to that for
$L_0$. 
\section{Results of Renormalization Group calculations}
We construct the phase diagram primarily by studying the RG flow pattern. We always start our iteration with 
$\epsilon_{A}=\epsilon_{B}=0$ and $U_A/t, U_B/t \ge 0$.
A transition point can be easily identified by looking into the RG flow diagram in the effective parameter space $\{U_{B}/t,|(\epsilon_{A}-\epsilon_{B})/t|\}$ at a given value of  $U_{A}/t$. For each value of $U_A/t$, there exists a
point on the $U_B/t$ axis (with $0 \le U_B/t \le U_A/t$) which behaves like a ``point of repulsion'' between the flow lines (fig. 4).

\begin{figure}
\includegraphics*{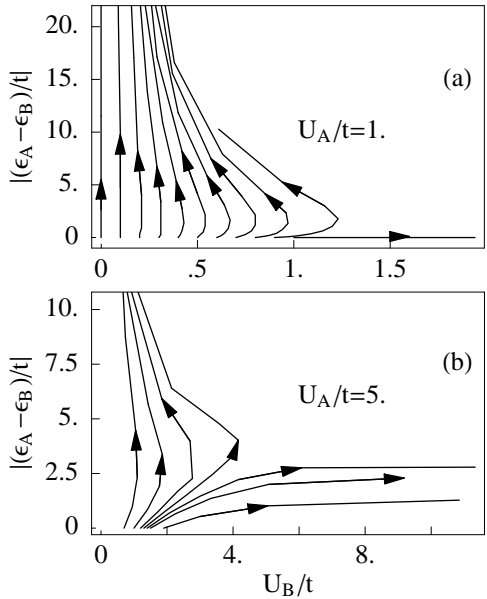}
\caption{ RG flow diagram in an effective parameter space $ U_B/t - |(\epsilon_A - \epsilon_B)/t|$ for (a) $U_A/t=1.0$, and (b) $U_A/t=5.0$. Point of repulsion of flow lines on the $U_B/t$ axis identifies a transition point between two types of phases.}
\end{figure}

Starting from any point on its left the RG flow tends to go to the fixed point $\{0,\infty\}$, indicating a charge ordered 
phase, while any point on its right flows to $\{\infty,0\}$  indicating a spin density wave. A plot of these transition
points on the $U_A/t - U_B/t$ plane shows the phase boundaries, which can also be viewed as ``lines of repulsion'' of the 
flow diagram projected on the $U_A/t - U_B/t$ plane (fig. 5). 

\begin{figure}
\includegraphics*{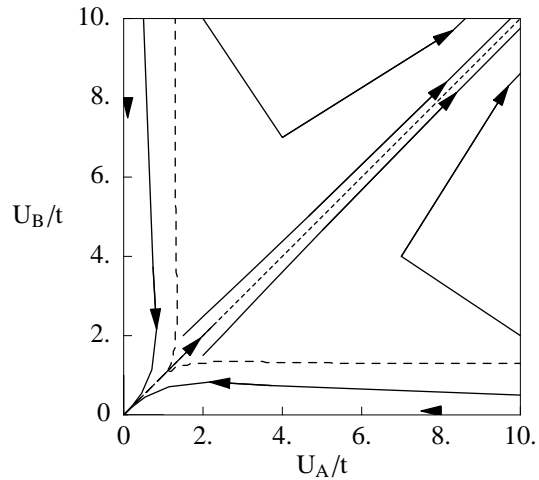}
\caption{RG flow diagram as projected on the $U_A/t - U_B/t$ plane together with the phase boundaries (dashed lines) which seem to be the ``lines of repulsion'' of the flow lines. The ``line of homogeneity'' is shown by the dotted line.}
\end{figure}

At this point it is interesting to find out the energy gap in
the spectrum, $\Delta_{RG}$, from the present RG calculation. For the present case we find that the Hamiltonian always flows
under the RG iterations to fixed points corresponding to the ``atomic limit'' ($t \rightarrow 0$) of (5) with parameters
$\epsilon_A^{(\infty)}$, $\epsilon_B^{(\infty)}$, and $U_A^{(\infty)} = U_B^{(\infty)} = U^{(\infty)}$, where the 
superscript $(\infty)$ refers to the converged values of the corresponding parameters in (5). Therefore, the gap can be calculated from $\Delta_{RG}= |(|\epsilon_A^{(\infty)} - \epsilon_B^{(\infty)}|- U^{(\infty)})|$. We find that at the 
transition points the energy gap in the spectrum vanishes. 

\begin{figure}
\includegraphics*{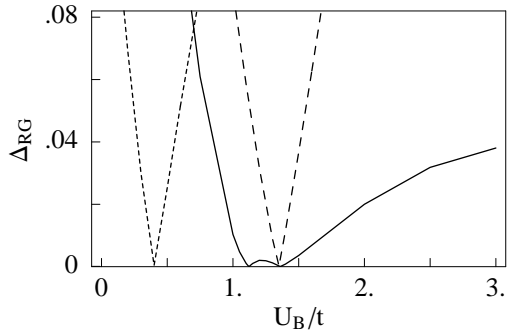}
\caption{Plot of the energy gap $\Delta_{RG}$, as found from the RG calculation, against $U_B/t$. The dotted, solid, and the dashed lines correspond to $U_{A}=0.4,~1.2$, and $2.5$ respectively (scale of energy is chosen by
setting $t=1.0$).}
\end{figure}

In fig. 6 we plot the energy gap $\Delta_{RG}$ as a 
function of $U_B/t$ for different values of 
$U_A/t$. It is interesting to note that the zeroes of $\Delta_{RG}$  occur at values of $U_A$ and
$U_B$ which are very close to the corresponding values obtained at zeroes of $\Delta_{HFA}$. 

\begin{figure}
\includegraphics*{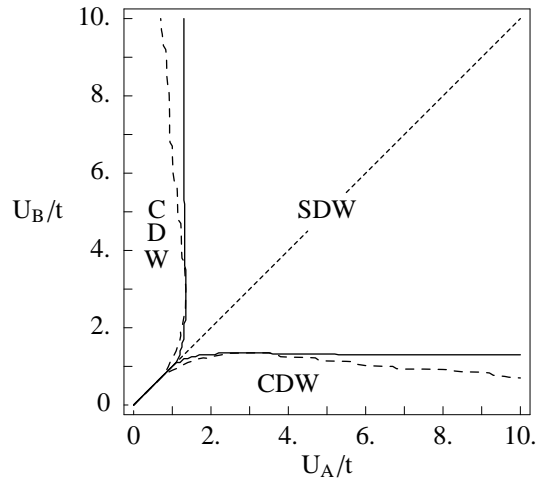}
\caption{Phase diagram obtained from the RG (solid line) superimposed on the HFA phase diagram (dashed line) in the $U_A/t - U_B/t$ plane. The ``line of homogeneity'' is shown by the dotted line.}
\end{figure}

Therefore, the phase diagram 
obtained in the
RG calculation agrees fairly well with that obtained in the HFA calculation 
(fig. 7). Departures are appreciable only at the strong
coupling limit; it can be easily understood because in this limit the error due to the truncation of basis becomes most 
serious in the present RG scheme for a model having two types of cells \cite{Gupta1}. The energy scales used for truncation 
of basis in the two types of cells now become appreciably different; the RG result, therefore, may not be highly reliable 
in this sector. However, the agreement of the RG and the HFA results from the weak- to the intermediate coupling regime 
really indicates that these results are very much reliable.

\begin{figure}
\includegraphics*{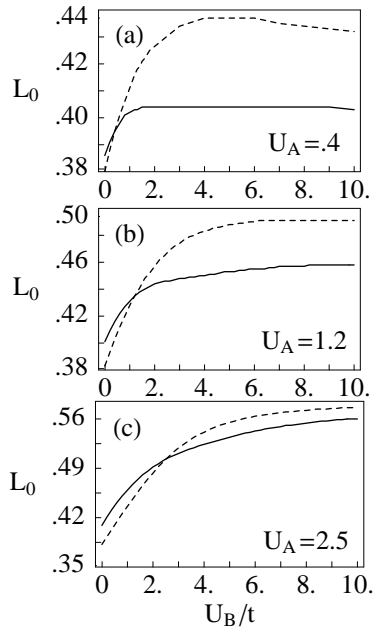}
\caption{Plot of the local moments $L_0$ as functions of $U_B/t$ for (a) $U_A =0.4$, (b) $U_A = 1.2$, and (c) $U_A=2.5$ (scale of energy is chosen by
setting $t=1.0$). The solid line shows the data for the $ABA$ block, and the dashed line shows the data for the $BAB$ block. The curves cross over each other at the ``point of homogeneity''.}
\end{figure}

In fig. 8 we plot the local moments at two types of sites as functions of $U_B/t$ for different values of $U_A/t$. It turns out that $L_0$ is always higher at sites with larger values of Coulomb correlation. This is in agreement with a previous observation in \cite{Santos1}. $L_0$ at $A$ type site equals that at $B$ type site only at the point of homogeneity i.e. at $U_A =U_B$. It is rather interesting to note that the values of $L_0$ is slightly larger than $\frac{3}{8}$ even in the parameter space where, according to the RG flow pattern (and also from the HFA), the system develops a CDW instability. This shows that this correlation-driven CDW phase is dominated by short length-scale fluctuations that are suppressed in a CDW phase generated by a periodic modulation in the site potentials alone \cite{Gupta1}.

\begin{figure}
\includegraphics*{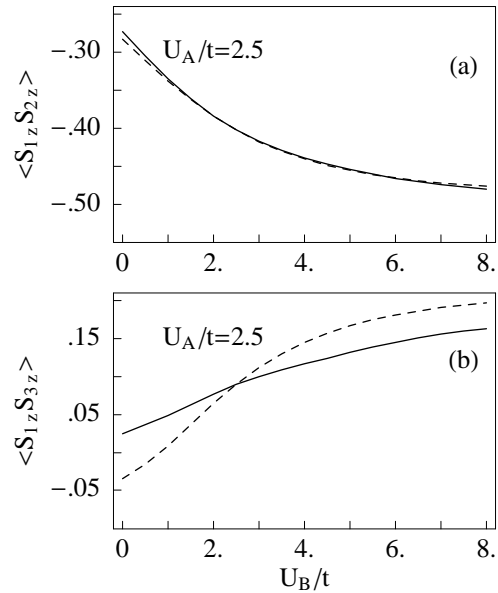}
\caption{Plot of the spin-spin correlation functions as functions of $U_B/t$ $ (U_A/t=2.5)$ for (a) nn correlation, and (b) nnn correlation. The solid line shows the data for the $ABA$ block, and the dashed line shows the data for the $BAB$ block.}
\end{figure}

We also calculate the nn and nnn spin-spin correlation functions with reference to $ABA$ and $BAB$ type cells. Plotted as 
functions of $U_B/t$ these (fig. 9) reveal some interesting points. The negative value of $\langle S_{1z} S_{2z}\rangle$ 
indicates a nearest neighbor antiferromagnetic alignment, which is increasing from the CDW to the SDW regime, as one should 
expect on physical grounds. That this correlation is not zero even within the CDW phase is really a reflection of the 
short-range fluctuations that we have just mentioned. However, the nn correlation has slightly different values in $ABA$ and 
$BAB$ cells. This difference, which is much more pronounced at higher values of $|U_A - U_B|$, is possibly related to the 
finite size effect of the RG. This effect is suppressed at weak coupling and near the ``homogeneous'' point in the SDW region.
On the other hand, the nnn correlation has opposite signs for the $ABA$ and the $BAB$ type cells in a region corresponding 
to the 
charge density instability. This indicates a frustration that suppresses the large distance antiferromagnetic correlation. This
frustration persists even within the SDW phase as it is evident from the widely different values of the nnn correlation in 
$ABA$ and $BAB$
type cells.  Such an effect has already been anticipated in a previous work \cite{Santos1}. It is 
to be noted here that this effect gradually reduces as $U_B \rightarrow U_A$. In this 
respect the ``SDW phase'' of the half-filled AHM is behaving in a different way than the SDW instability found in a 
half-filled Hubbard model (the ``homogeneous limit'' of the present model (1)).
\section{Study of Magnetoconductance}
Now we study the nature of the ground state of the superlattice structure in presence of a magnetic field. The magnetic field penetrating the ring will interact with the moments of the electrons. There will be an additive Zeeman term of the form $\mu_{\sigma} H'$ in the Hamiltonian (1), where $\mu_{\sigma}$ is the moment of the spin $\sigma$ and $H'$ is the penetrating magnetic field. The Hamiltonian now becomes 
\begin{eqnarray}
&&H= H'\sum_{i,\sigma}\mu_{\sigma} n_{i,\sigma}+t\sum_{i,\sigma}(c_{i,\sigma}^{\dag}c_{i+1,\sigma}+h.c.)\nonumber\\
&&+U_{A}\sum_{i\in \cal{A}}n_{i,\uparrow}n_{i,\downarrow}+U_{B}\sum_{i\in \cal{B}}n_{i,\uparrow}n_{i,\downarrow}~.
\end{eqnarray}

\noindent
We follow the same HFA to decouple the Hamiltonian. Since we have found regions of the parameter space where the HFA results and the RG results match very well, we can rely on the HFA results in these regions. The Hamiltonian corresponding to the 
up (down) spin electrons will generate the up (down) spin band. These two bands are not degenerate because the spin 
reversal symmetry is now broken. Therefore, at half-filling, the number of up and down spin electrons will be different and a 
net moment will be generated. 

For non-zero values of $U_{A} (>0$) and  $U_{B}=0$, the system is charge ordered at half filling in absence of a magnetic 
field. This is an insulating phase with a gap between the upper and lower bands; at half filling the lower bands for both spin 
species are totally filled. When a small magnetic field is turned on, the up and the down spin bands shift in the opposite 
directions in energy scale. At a sufficiently large value of the magnetic field the upper band of the up spin electrons tend to get occupied at the cost of depopulating the lower band of the down spin electrons. Thus the ground state contains unequal 
number of up and down spins and the system, now having two partially filled bands, becomes conducting. If we go on increasing 
the magnetic field beyond a certain value, the said up spin band will be completely filled while the 
down spin band will be completely empty. The system is now spin polarized, leading again to an insulating phase. This can be 
easily checked by calculating the Drude weight \cite{Kohn} which measures the dc conductivity of the chain. Similar analysis 
can be made 
for $U_B \neq 0$ as well. Of course, in this case there arises a possibility of transition to a conducting phase from an SDW 
under the application of the field $H'$.
We have  calculated the Drude weight both for $U_B=0$ and $U_B =2$. To calculate the Drude weight, a
vanishingly small magnetic flux $\Phi$ (in units of flux quantum $\phi_{0}=hc/e$) is introduced. The ring encloses this flux, but the flux does not penetrate the ring. Now the hopping term is modified by a phase factor. The Hamiltonian becomes
\begin{eqnarray}
H&=& H'\sum_{i,\sigma}\mu_{\sigma} n_{i,\sigma}\nonumber\\
&&+t\sum_{i,\sigma}(c_{i,\sigma}^{\dag}c_{i+1,\sigma} e^{2\pi i\phi}+h.c.)\nonumber\\
&&+U_{A}\sum_{i\in \cal{A}}n_{i,\uparrow}n_{i,\downarrow}+U_{B}\sum_{i\in \cal{B}}n_{i,\uparrow}n_{i,\downarrow}
\end{eqnarray}

\noindent
where $\phi=\Phi/N$, $N$ being the number of sites in the ring.  The Drude weight is calculated from the formula \cite{Gupta1, Gupta2}
\begin{equation}
D=\frac{N}{4\pi^{2}}\left[\frac{\partial^{2}E(\Phi)}{\partial\Phi^{2}}\right]_{\Phi=0}
\end{equation}

\noindent
where $E(\Phi)$ is the ground state energy of (18) calculated within the HFA.

\begin{figure}
\includegraphics*{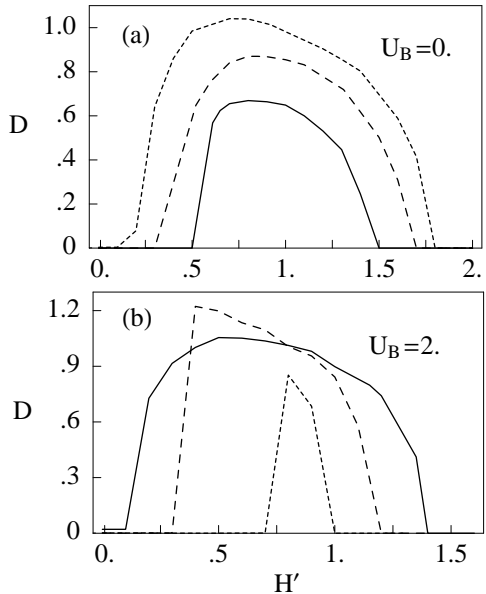}
\caption{Plot of the Drude weight $D$ vs. the field $H'$ for (a) $U_B=0.$ and (b) $U_B=2.0$. The dotted, dashed and the solid lines correspond to $U_{A}=1.0$, $2.0$, and $ 4.0$ respectively in (a), and to $U_{A}=3.0$, $2.0$, and $ 0.9$ respectively in (b) (scale of energy is chosen by
setting $t=1.0$).}
\end{figure}

We have plotted the Drude weight $D$ against the penetrating magnetic field $H'$ (fig. 10) with $N= 300$ for  fixed set of
values of $U_{A}$ and $U_{B}$. The curves clearly show the transition between conducting and 
insulating phases. For lower and higher values of $H'$ the Drude weight is zero i.e. the system is insulating. But for moderate
values of $H'$, (depending on the value of on-site Coulomb repulsion energy) the Drude weight is quite large, implying a 
conducting phase. For higher values of $U_{A}$ the conducting region becomes narrower. For different combinations of $U_{A}$ 
and $U_{B}$ values, we have plotted (fig. 11) the Drude weight $D$ against system size $N$. For the values of the penetrating 
magnetic field $H'$, for which the upper band of the up-spins starts to depopulate and the lower down-spin band gets partially
populated, the system goes to a conducting phase. In fact, the Drude weight remains unaltered as we increase the system 
size.

\begin{figure}
\includegraphics*{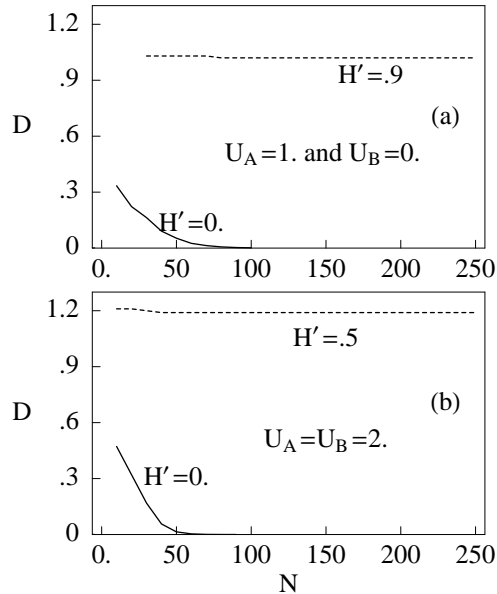}
\caption{Plot of the Drude weight $D$ vs. system size $N$ for (a) $U_A=1.0$ and $U_B=0$, and (b) $U_A =U_B=2.0$ (scale of energy is chosen by
setting $t=1.0$). Cases with $H'=0$ (solid line) and $H'>H'_c$ (dashed line) are shown where $H'_c$ is the value of $H'$ for which the system enters a conducting phase from an insulating one.}
\end{figure}

For $H'=0$, the present system is found to be either in a CDW  or in an SDW phase (fig. 1) which is insulating. Therefore,
the conductivity of such a system of macroscopic size must be vanishing. Thus  in this region, the Drude weight  falls sharply 
with increasing system size. Similar things happen for small values of $H'$ for which the system is yet to pass on to the conducting 
phase. To identify the nature of the 
insulating phases directly, we study the spin and charge density parameters, $s$ and $c$ respectively, together with the magnetization
\begin{eqnarray}
m = \frac{1}{2} \langle n_{B,\uparrow}-n_{B,\downarrow}+n_{A,\uparrow}-n_{A,\downarrow}\rangle\nonumber~.
\end{eqnarray}

\begin{figure}
\includegraphics*{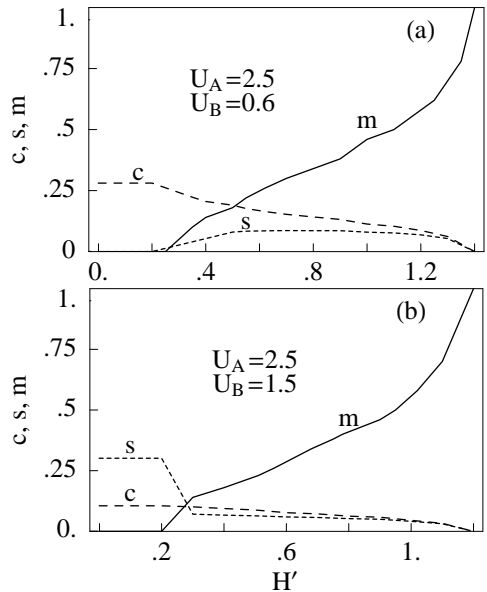}
\caption{Plot of the order parameters $c, s$, and $m$ against the field $H'$. The solid line shows magnetization, the dotted line shows spin order parameter, and the dashed line shows charge order parameter for (a) $U_B=0.6$ and (b) $U_B  =1.5$ while $U_A = 2.5$ (scale of energy is chosen by
setting $t=1.0$) and $N =100$.}
\end{figure}

Fig. 12 shows  the plots of $c, s$, and $m$ as functions of $H'$ for different values of $U_A$ and $U_B$. For $U_A >> U_B$ 
we find that the increase in $H'$ drives the system  from a charge ordered phase (large $c$) to a phase where $c, s,$ and 
$m$ are nearly comparable; this indicates the existence of a conducting state due to the absence of any kind of long range ordering. 
We have already noted in fig. 10 that in this region $D$ is non-zero which is consistent with the above observation. Further increase in $H'$ 
would completely depopulate the down-spin bands and consequently the up-spin bands would be filled up. This
induces a sharp rise in $m$ (together with a vanishing $D$) indicating a transition into an insulating spin polarized phase. For $U_B \sim U_A (\sim t)$, however, the initial transition is from a spin ordered phase to a metallic one.

\begin{figure}
\includegraphics*{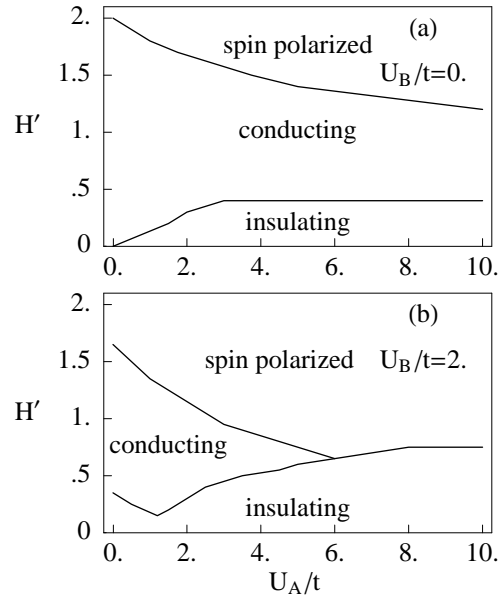}
\caption{The HFA phase diagram of the AHM in presence of a Zeeman field $H'$ for $N= 100$ and for (a) $U_B/t=0$, and 
(b) $U_B/t=2$}
\end{figure}

Thus we can 
construct 
the phase diagram (fig. 13) of the AHM in presence of a Zeeman field which shows the possibility of having an enhanced
magnetoconductance for moderate values of the repulsive interactions. In case of very large Coulomb interaction, however, 
no conducting phase appears, and a direct transition from an insulating SDW phase to a spin polarized phase is observed.
\section{Conclusion}
Summarizing, we have investigated the one dimensional half-filled alternating Hubbard super lattice structure at zero 
temperature using the Hartree-Fock approximation and a real space renormalization group technique. Although both the methods 
are approximate ones, they are complementary to one other. The agreement between the  phase diagrams obtained from these two 
different methods shows that the results are rather reliable. We obtained two types of phases, dominated by charge and spin 
density wave instabilities respectively, depending on the $U_{A}$ and $U_{B}$ values. The $U_{A}-U_{B}$ phase diagram shows two transition lines, indicating a centrally located antiferromagnetic region and two charge density wave  regions near the axes.

The system is
insulating. It may be noted at this point that the possibility of such transitions was not explored in the previous works 
on this model; most of these works concentrated only on a limited region of the parameter space, e.g. $U_A > 0$ and $U_B=0$,
and also there were some severe restrictions on the system size. Apart from the phase transitions, it is interesting to note that there appears a spatial 
modulation of the local moments dictated by the inhomogeneity of the correlation parameters. The RG results also indicate that
the ordering is not full grown as compared to the homogeneous limit because of underlying frustration of spin-spin correlation.
These results are in agreement with a previous finding \cite{Santos1}. Also
we have studied the ground state properties, including the Drude weight,  of the superlattice  in presence of a magnetic field
$(H')$. The $U_A-H'$ (for fixed $U_B$) phase diagram shows that the system becomes conducting for an intermediate range of values of $H'$.  In the phase
diagram, we obtained insulating phases for lower and higher values of $H'$. The width of the conducting region depends on the 
values of on-site Coulomb repulsion energies of the super lattice structure. 
It seems interesting to explore this model further, especially at finite temperatures and for cases away from  half-filling.
Also the effect of varying ``spacer thickness'' \cite{Santos1, Santos2, Santos3} for this regime of correlation parameters may yield
some interesting observations.

\begin{acknowledgments}
Two of the authors (JC and BB) sincerely acknowledge a fruitful discussion with Dr. Shreekantha Sil. 
\end{acknowledgments}

\end{document}